\pdfoutput=1
\documentclass[12pt,a4paper]{article}
\usepackage{ifthen} % for conditional statements
\newboolean{pdflatex}
\setboolean{pdflatex}{true} % False for eps figures 

\newboolean{articletitles}
\setboolean{articletitles}{true} % False removes titles in references

\newboolean{uprightparticles}
\setboolean{uprightparticles}{false} %True for upright particle symbols

\def\paperauthors{LHCb collaboration} % Leave as is for PAPER, CONF and FIGURE
\def\paperasciititle{Measurement of phi meson production in fixed-target pNe collisions at sqrt(s)NN = 68.5 GeV} % Set ASCII title here !! MAKE sure it's only ASCII characters !! 
\def\papertitle{Measurement of $\phi(1020)$ meson production in fixed-target $\textit{p}$Ne collisions at~\sqsnn~=~68.5\gev} % Latex formatted title
\def\paperkeywords{{High Energy Physics}, {LHCb}} % Comma separated list
\def\papercopyright{\the\year\ CERN for the benefit of the LHCb collaboration} % new since 9/Apr/2018
\def\paperlicence{CC BY 4.0 licence}
\def\paperlicenceurl{https://creativecommons.org/licenses/by/4.0/}

\newif\ifEnableSectionTOCLinks
\EnableSectionTOCLinksfalse % deactivated

\usepackage[top=1in, bottom=1.25in, left=1in, right=1in]{geometry}

\usepackage{microtype}
\usepackage{lineno}  % for line numbering during review
\usepackage{xspace} % To avoid problems with missing or double spaces after
\usepackage{caption} %these three command get the figure and table captions automatically small

\usepackage{graphicx}  % to include figures (can also use other packages)
\usepackage{color}
\usepackage{colortbl}
\graphicspath{{./figs/}} % Make Latex search fig subdir for figures
\usepackage{amsmath} % Adds a large collection of math symbols
\usepackage{amssymb}
\usepackage{amsfonts}
\usepackage{upgreek} % Adds in support for greek letters in roman typeset

\newcommand*\patchAmsMathEnvironmentForLineno[1]{%
\expandafter\let\csname old#1\expandafter\endcsname\csname #1\endcsname
\expandafter\let\csname oldend#1\expandafter\endcsname\csname
end#1\endcsname
 \renewenvironment{#1}%
   {\linenomath\csname old#1\endcsname}%
   {\csname oldend#1\endcsname\endlinenomath}%
}
\newcommand*\patchBothAmsMathEnvironmentsForLineno[1]{%
  \patchAmsMathEnvironmentForLineno{#1}%
  \patchAmsMathEnvironmentForLineno{#1*}%
}
\AtBeginDocument{%
\patchBothAmsMathEnvironmentsForLineno{equation}%
\patchBothAmsMathEnvironmentsForLineno{align}%
\patchBothAmsMathEnvironmentsForLineno{flalign}%
\patchBothAmsMathEnvironmentsForLineno{alignat}%
\patchBothAmsMathEnvironmentsForLineno{gather}%
\patchBothAmsMathEnvironmentsForLineno{multline}%
\patchBothAmsMathEnvironmentsForLineno{eqnarray}%
}

\usepackage{hyperref}
\usepackage{hyperxmp}
\usepackage{hyperref}[pdftex,
            pdfauthor={\paperauthors},
            pdftitle={\paperasciititle},
            pdfkeywords={\paperkeywords},
            pdfcopyright={Copyright (C) \papercopyright},
            pdflicenseurl={\paperlicenceurl}]%{hyperref}
\usepackage[colorinlistoftodos,textsize=scriptsize]{todonotes}

\usepackage[bottom,flushmargin,hang,multiple]{footmisc}

\usepackage[all]{hypcap} % Internal hyperlinks to floats.

\usepackage{xspace} 
\usepackage{upgreek}

\def\lhcb   {\mbox{LHCb}\xspace}

\def\MagUp {\mbox{\em Mag\kern -0.05em Up}\xspace}

\ifthenelse{\boolean{uprightparticles}}%
{

 \def\Pmu         {\ensuremath{\upmu}\xspace}

 \def\Ppi         {\ensuremath{\uppi}\xspace}

 \def\PDelta      {\ensuremath{\Delta}\xspace}                 
 \def\PXi         {\ensuremath{\Xi}\xspace}                 
 \def\PLambda     {\ensuremath{\Lambda}\xspace}                 
 \def\PSigma      {\ensuremath{\Sigma}\xspace}                 
 \def\POmega      {\ensuremath{\Omega}\xspace}                 
 \def\PUpsilon    {\ensuremath{\Upsilon}\xspace}
 \let\oldPi\Pi
 \def\PPi         {\ensuremath{\oldPi}\xspace}

 \def\PB      {\ensuremath{\mathrm{B}}\xspace}                 
 \def\PD      {\ensuremath{\mathrm{D}}\xspace}                 

 \def\PK      {\ensuremath{\mathrm{K}}\xspace}                 

 \def\Pc      {\ensuremath{\mathrm{c}}\xspace}                 
                  
 \def\Pe      {\ensuremath{\mathrm{e}}\xspace}                 
 \def\Pp      {\ensuremath{\mathrm{p}}\xspace}                 

 \def\Ps      {\ensuremath{\mathrm{s}}\xspace}

 \def\thebaroffset{0.0em}
}
{

 \def\Pmu         {\ensuremath{\mu}\xspace}

 \def\Ppi         {\ensuremath{\pi}\xspace}

 \mathchardef\PDelta="7101
 \mathchardef\PXi="7104
 \mathchardef\PLambda="7103
 \mathchardef\PSigma="7106
 \mathchardef\POmega="710A
 \mathchardef\PUpsilon="7107
 \mathchardef\PPi="7105
 \def\PB      {\ensuremath{B}\xspace}                 
 \def\PD      {\ensuremath{D}\xspace}                 

 \def\PK      {\ensuremath{K}\xspace}                 

 \def\Pc      {\ensuremath{c}\xspace}                 
                  
 \def\Pe      {\ensuremath{e}\xspace}                 
 \def\Pp      {\ensuremath{p}\xspace}                 

 \def\Ps      {\ensuremath{s}\xspace}

 \def\thebaroffset{0.18em}
}
\newcommand{\offsetoverline}[2][\thebaroffset]{\kern #1\overline{\kern -#1 #2}}%

\makeatletter
\ifcase \@ptsize \relax% 10pt
  \newcommand{\miniscule}{\@setfontsize\miniscule{4}{5}}% \tiny: 5/6
\or% 11pt
  \newcommand{\miniscule}{\@setfontsize\miniscule{5}{6}}% \tiny: 6/7
\or% 12pt
  \newcommand{\miniscule}{\@setfontsize\miniscule{5}{6}}% \tiny: 6/7
\fi
\makeatother

\DeclareRobustCommand{\optbar}[1]{\shortstack{{\miniscule (\rule[.5ex]{1.25em}{.18mm})}
  \\ [-.7ex] $#1$}}

\def\electron   {{\ensuremath{\Pe}}\xspace}
\def\muon       {{\ensuremath{\Pmu}}\xspace}

\def\squark    {{\ensuremath{\Ps}}\xspace}

\def\cquark    {{\ensuremath{\Pc}}\xspace}

\def\pion   {{\ensuremath{\Ppi}}\xspace}

\def\pip    {{\ensuremath{\pion^+}}\xspace}

\def\kaon    {{\ensuremath{\PK}}\xspace}
\def\KorKbar {\kern \thebaroffset\optbar{\kern -\thebaroffset \PK}{}\xspace}

\def\Kp      {{\ensuremath{\kaon^+}}\xspace}
\def\Km      {{\ensuremath{\kaon^-}}\xspace}

\def\D       {{\ensuremath{\PD}}\xspace}

\def\DorDbar {\kern \thebaroffset\optbar{\kern -\thebaroffset \PD}\xspace}

\def\Dp      {{\ensuremath{\D^+}}\xspace}
\def\Dm      {{\ensuremath{\D^-}}\xspace}

\def\DpDm    {\ensuremath{\Dp {\kern -0.16em \Dm}}\xspace}

\def\B       {{\ensuremath{\PB}}\xspace}

\def\BorBbar {\kern \thebaroffset\optbar{\kern -\thebaroffset \PB}\xspace}

\def\Bd      {{\ensuremath{\B^0}}\xspace}

\def\BdorBdbar {\kern \thebaroffset\optbar{\kern -\thebaroffset \Bd}\xspace}
\def\Bu      {{\ensuremath{\B^+}}\xspace}

\def\Bp      {{\ensuremath{\Bu}}\xspace}

\def\Bs      {{\ensuremath{\B^0_\squark}}\xspace}

\def\BsorBsbar {\kern \thebaroffset\optbar{\kern -\thebaroffset \Bs}\xspace}
\def\Bc      {{\ensuremath{\B_\cquark^+}}\xspace}

\def\Y#1S{\ensuremath{\PUpsilon{(#1S)}}\xspace}

\def\proton      {{\ensuremath{\Pp}}\xspace}

\def\LorLbar     {\kern \thebaroffset\optbar{\kern -\thebaroffset \PLambda}\xspace}

\def\BF         {{\ensuremath{\mathcal{B}}}\xspace}

\def\to                 {\ensuremath{\rightarrow}\xspace}

\newcommand{\mBs}{{\ensuremath{m_{\Bs}}}\xspace}

\def\CP                {{\ensuremath{C\!P}}\xspace}

\def\AT#1     {\ensuremath{A_{\mathrm{T}}^{#1}}\xspace}           % 2

\def\C#1      {\ensuremath{\mathcal{C}_{#1}}\xspace}                       % 9
\def\Cp#1     {\ensuremath{\mathcal{C}_{#1}^{'}}\xspace}                    % 7
\def\Ceff#1   {\ensuremath{\mathcal{C}_{#1}^{\mathrm{(eff)}}}\xspace}        % 9  
\def\Cpeff#1  {\ensuremath{\mathcal{C}_{#1}^{'\mathrm{(eff)}}}\xspace}       % 7
\def\Ope#1    {\ensuremath{\mathcal{O}_{#1}}\xspace}                       % 2
\def\Opep#1   {\ensuremath{\mathcal{O}_{#1}^{'}}\xspace}                    % 7

\newcommand{\nospaceunit}[1]{\ensuremath{\text{#1}}}       
\newcommand{\aunit}[1]{\ensuremath{\text{\,#1}}}       
\newcommand{\tev}{\aunit{Te\kern -0.1em V}\xspace}
\newcommand{\gev}{\aunit{Ge\kern -0.1em V}\xspace}
\newcommand{\mev}{\aunit{Me\kern -0.1em V}\xspace}
\newcommand{\kev}{\aunit{ke\kern -0.1em V}\xspace}
\newcommand{\ev}{\aunit{e\kern -0.1em V}\xspace}
 
\newcommand{\mevc}{\ensuremath{\aunit{Me\kern -0.1em V\!/}c}\xspace}
\newcommand{\gevc}{\ensuremath{\aunit{Ge\kern -0.1em V\!/}c}\xspace}
\newcommand{\mevcc}{\ensuremath{\aunit{Me\kern -0.1em V\!/}c^2}\xspace}
\newcommand{\gevcc}{\ensuremath{\aunit{Ge\kern -0.1em V\!/}c^2}\xspace}
\def\mm   {\aunit{mm}\xspace}

\def\mum  {\ensuremath{\,\upmu\nospaceunit{m}}\xspace}

\def\mub{\ensuremath{\,\upmu\nospaceunit{b}}\xspace}
\def\nb {\aunit{nb}\xspace}
\def\invnb {\ensuremath{\nb^{-1}}\xspace}

\def\deriv {\ensuremath{\mathrm{d}}}

\def\gsim{{~\raise.15em\hbox{$>$}\kern-.85em
          \lower.35em\hbox{$\sim$}~}\xspace}
\def\lsim{{~\raise.15em\hbox{$<$}\kern-.85em
          \lower.35em\hbox{$\sim$}~}\xspace}

\def\sPlot{\mbox{\em sPlot}\xspace}

\def\sqsnn {\ensuremath{\protect\sqrt{s_{\scriptscriptstyle\text{NN}}}}\xspace}
\def\pt         {\ensuremath{p_{\mathrm{T}}}\xspace}

\def\dirac      {\mbox{\textsc{Dirac}}\xspace}
\def\evtgen     {\mbox{\textsc{EvtGen}}\xspace}

\def\geant      {\mbox{\textsc{Geant4}}\xspace}

\def\photos     {\mbox{\textsc{Photos}}\xspace}

\def\pythia     {\mbox{\textsc{Pythia}}\xspace}

\def\tensorflow {\mbox{\textsc{TensorFlow}}\xspace}

\def\tell1  {TELL1\xspace}
\def\ukl1   {UKL1\xspace}

\newcommand{\eg}{\mbox{\itshape e.g.}\xspace}

\newcommand{\lhcborcid}[1]{\href{https://orcid.org/#1}{\hspace*{0.1em}\raisebox{-0.45ex}{\includegraphics[width=1em]{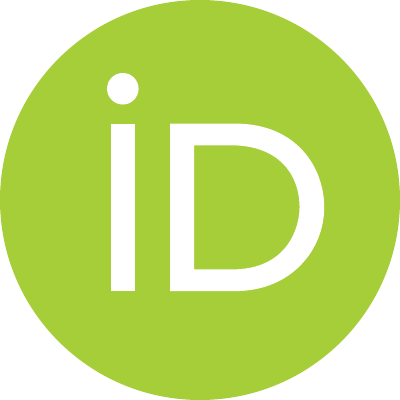}}}}

\hypersetup{
  colorlinks   = true, %Colours links instead of ugly boxes
  urlcolor     = blue, %Colour for external hyperlinks
  linkcolor    = blue, %Colour of internal links
  citecolor    = red   %Colour of citations
}

\ifEnableSectionTOCLinks
    \usepackage[explicit]{titlesec} % to change headings
    
    \let\oldcontentsline\contentsline
    \renewcommand

    \titleformat{\section}{\normalfont\Large\bf}{\hyperlink{tocsection.\thesection}{{\thesection} \parbox[t]{\dimexpr\textwidth-1pc}{#1}}}{1pc}{}

    \titleformat{\subsection}{\normalfont\bf}{\hyperlink{tocsubsection.\thesubsection}{{\thesubsection} \parbox[t]{\dimexpr\textwidth-1pc}{#1}}}{1pc}{}

    \titleformat{name=\section,numberless}[display]{}{}{0pt}{\normalfont\Huge\bfseries #1}
\fi

\usepackage{cite} % Allows for ranges in citations
\usepackage{mciteplus}
\usepackage{longtable} % only for template; not usually to be used in PAPERs
\usepackage{comment}
\usepackage{adjustbox}

\begin{document}

%%%%%%%%%%%%%%%%%%%%%%%%%
%%%%% Title     %%%%%%%%%
%%%%%%%%%%%%%%%%%%%%%%%%%
\renewcommand{\thefootnote}{\fnsymbol{footnote}}
\setcounter{footnote}{1}

% %%%%%%% CHOOSE TITLE PAGE--------
%\onecolumn
%\input{title-LHCb-INT}
%\input{title-LHCb-ANA}
%\input{title-LHCb-CONF}
%\input{title-LHCb-FIGURE}
% ===============================================================================
% Purpose: LHCb-PAPER journal paper title page template
% Author: 
% Created on: 2010-09-25
% ===============================================================================

%%%%%%%%%%%%%%%%%%%%%%%%%
%%%%%  TITLE PAGE  %%%%%%
%%%%%%%%%%%%%%%%%%%%%%%%%
\begin{titlepage}
\pagenumbering{roman}

% Header ---------------------------------------------------
\vspace*{-1.5cm}
\centerline{\large EUROPEAN ORGANIZATION FOR NUCLEAR RESEARCH (CERN)}
\vspace*{1.5cm}
\noindent
\begin{tabular*}{\linewidth}{lc@{\extracolsep{\fill}}r@{\extracolsep{0pt}}}
\ifthenelse{\boolean{pdflatex}}% Logo format choice
{\vspace*{-1.5cm}\mbox{\!\!\!\includegraphics[width=.14\textwidth]{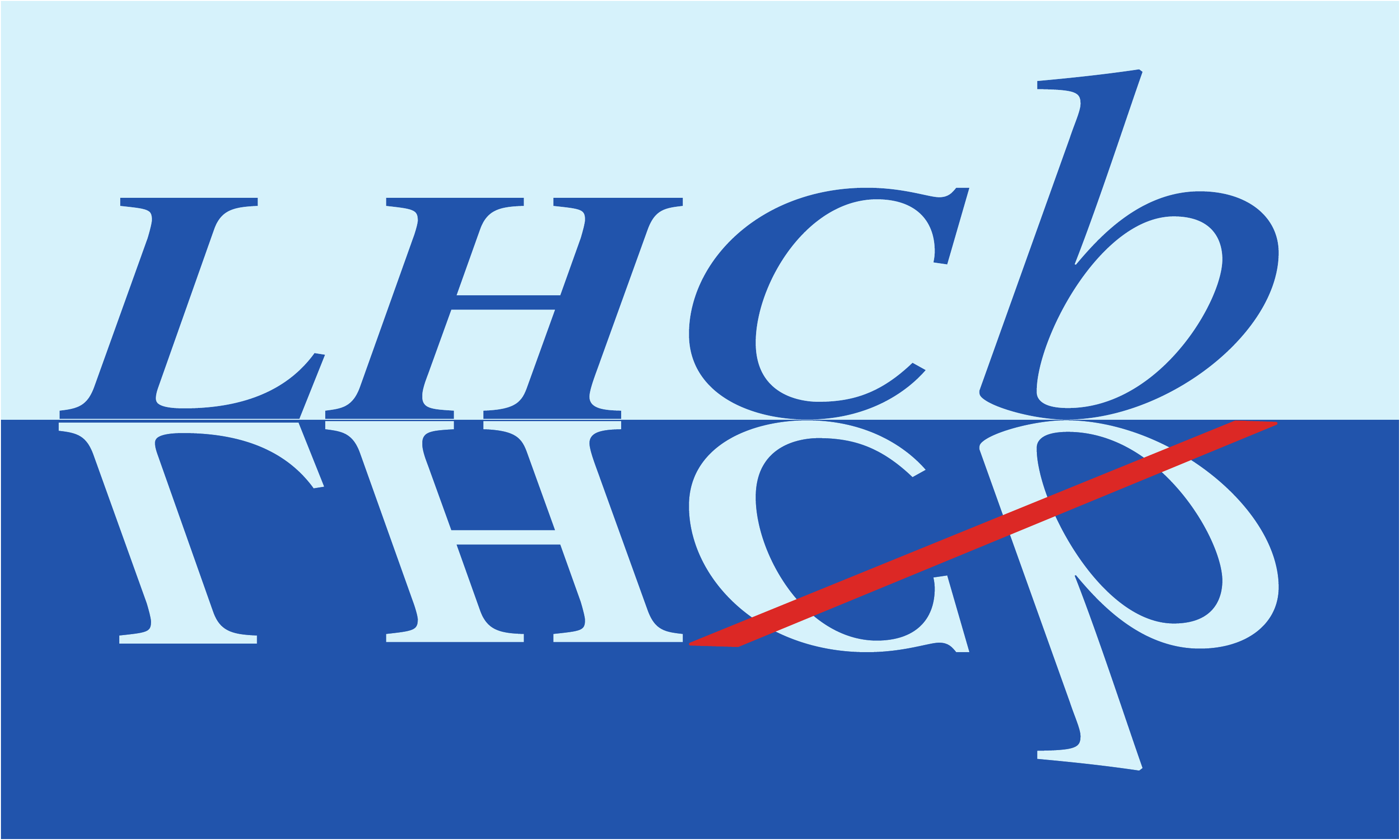}} & &}%
{\vspace*{-1.2cm}\mbox{\!\!\!\includegraphics[width=.12\textwidth]{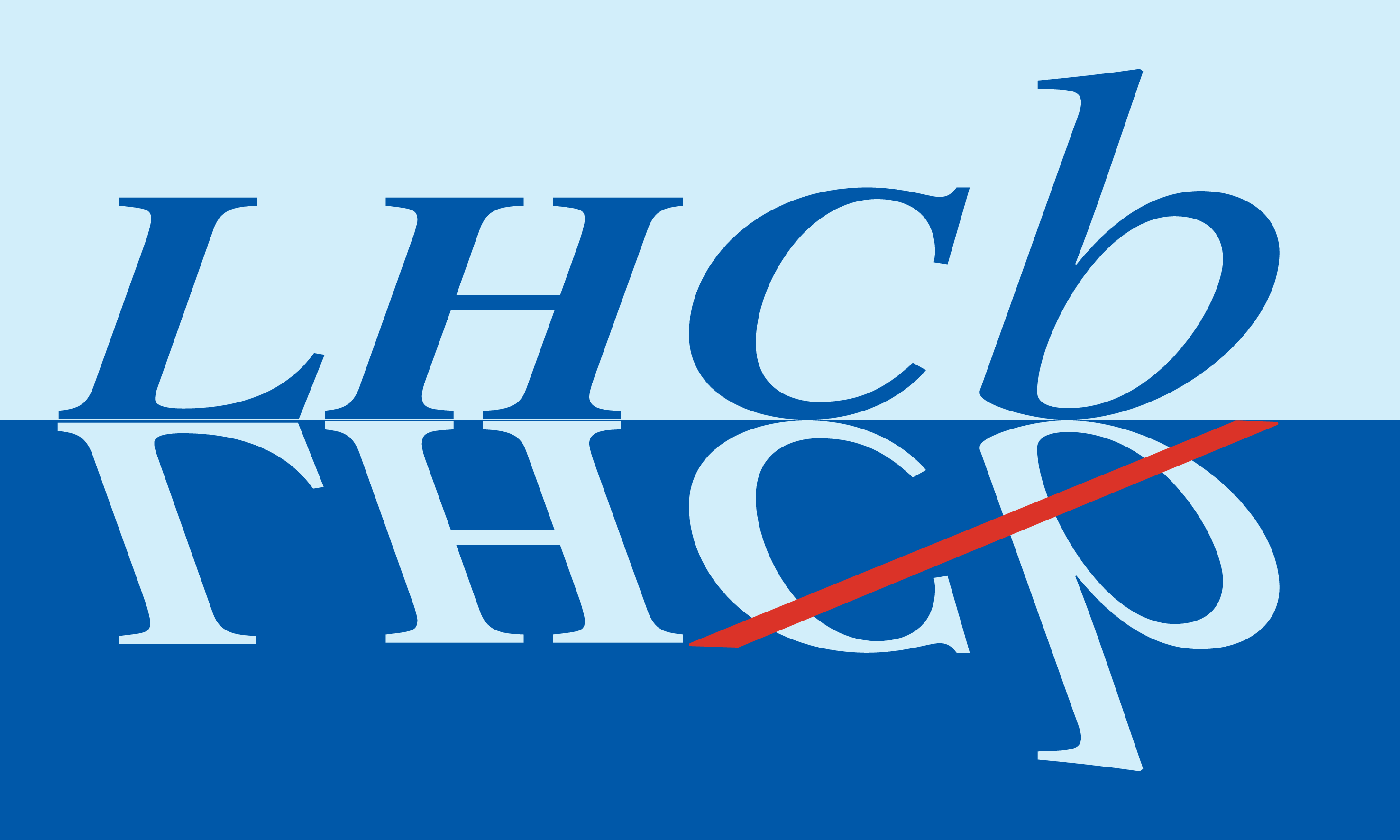}} & &}%
\\
 & & CERN-EP-2024-274 \\  % ID 
 & & LHCb-PAPER-2024-036 \\  % ID 
 & & March 20, 2025 \\ % Date - Can also hardwire e.g.: 23 March 2010
 & & \\
% not in paper \hline
\end{tabular*}

\vspace*{4.0cm}

% Title --------------------------------------------------
{\normalfont\bfseries\boldmath\huge
\begin{center}
% DO NOT EDIT HERE. Instead edit macro in main.tex to keep metadata correct
  \papertitle 
\end{center}
}

\vspace*{2.0cm}

% Authors -------------------------------------------------
\begin{center}
%In the footnote, replace 'paper' by 'Letter' in case of submission to PRL or PLB 
% Edit macro in main.tex to keep metadata correct
\paperauthors\footnote{Authors are listed at the end of this paper.}
\end{center}

\vspace{\fill}

% Abstract -----------------------------------------------
\begin{abstract}
  \noindent
The first measurement of $\phi(1020)$ meson production in fixed-target $p$Ne collisions at \mbox{$\sqsnn=68.5\gev$} is presented. 
The $\phi(1020)$ mesons are reconstructed in their $\Kp\Km$ decay in a data sample 
consisting of proton collisions on neon nuclei at rest, corresponding to an integrated luminosity of $21.7 \pm 1.4$\invnb, collected by the \lhcb detector at CERN. The $\phi(1020)$ production cross-section in the centre-of-mass rapidity range of \mbox{$-1.8<y^*<0$} and transverse momentum range of \mbox{$800<\pt <6500$\mevc} is found to be $\sigma=182.7\pm2.7~\text{(stat.)}\pm14.1~\text{(syst)}$\mub/nucleon.
A double-differential measurement of the cross-section is also provided in four regions of rapidity and six regions of transverse momentum of the $\phi(1020)$ meson and compared with the predictions from Pythia and EPOS4, which are found to underestimate the experimental values.
  
\end{abstract}

\vspace*{2.0cm}

\begin{center}
  Published in JHEP 03 (2025) 151
\end{center}

\vspace{\fill}

{\footnotesize 
% Edit macro in main.tex to keep metadata correct
\centerline{\copyright~\papercopyright. \href{\paperlicenceurl}{\paperlicence}.}}
\vspace*{2mm}

\end{titlepage}

%%%%%%%%%%%%%%%%%%%%%%%%%%%%%%%%
%%%%%  EOD OF TITLE PAGE  %%%%%%
%%%%%%%%%%%%%%%%%%%%%%%%%%%%%%%%

%  empty page follows the title page ----
\newpage
\setcounter{page}{2}
\mbox{~}
%\newpage
%
%% Author List ----------------------------
%%  You need to get a new author list!
%\input{LHCb_authorlist.tex}
%
%The author list for journal publications is provided by the Membership Committee shortly after 'approval to go to paper' has been given.
%%It will be made available on the page
%%\verb!http://www.physik.uzh.ch/~strauman/forMemCo/LHCb-PAPER-XXXX-XXX/! .
%It will be sent to you by email shortly after a paper number has beens assigned.
%The author list should be included already at first circulation, 
%to allow new members of the collaboration to verify whether they have been included correctly.
%Occasionally a misspelled name is corrected or associated institutions become full members.
%In that case, a new author list will be sent to you.
%In case line numbering doesn't work well after including the authorlist, try moving the \verb!\bigskip! after the last author to a separate line.
%
%
%The authorship for Conference Reports should be ``The LHCb
%  collaboration'', with a footnote giving the name(s) of the contact
%  author(s), but without the full list of collaboration names.

%\twocolumn
% %%%%%%%%%%%%% ---------

\renewcommand{\thefootnote}{\arabic{footnote}}
\setcounter{footnote}{0}

%%%%%%%%%%%%%%%%%%%%%%%%%%%%%%%%
%%%%%  Table of Content   %%%%%%
%%%%%%%%%%%%%%%%%%%%%%%%%%%%%%%%
%%%% Uncomment if desired
%\tableofcontents

\cleardoublepage

%%%%%%%%%%%%%%%%%%%%%%%%%
%%%%% Main text %%%%%%%%%
%%%%%%%%%%%%%%%%%%%%%%%%%

\pagestyle{plain} % restore page numbers for the main text
\setcounter{page}{1}
\pagenumbering{arabic}

%% Uncomment during review phase. 
%% Comment before a final submission.
%\linenumbers

%% This is the main body
%% It is useful to have a single file so comments are not missed in overleaf.
\section{Introduction}
\label{sec:Introduction}

The existence of a quark-gluon plasma (QGP), a state of matter where quarks and gluons are deconfined, is allowed by quantum chromodynamics (QCD) at high temperature and high baryon density~\cite{Shuryak:1980tp}. Experimental evidence of QGP formation has been found in large collision systems~\cite{STAR:2005gfr,PHENIX:2004vcz,ALICE:2022wpn},
while formation of QGP droplets has been proposed to describe some observables in collisions of small asymmetric systems (\eg proton-nucleus) at $\sqsnn=200$\gev~\cite{PHENIX:2018lia}. However, in proton-nucleus collisions, cold nuclear matter (CNM) effects are also present, such as modifications of the initial parton distribution functions in nuclei (nPDFs) due to shadowing~\cite{deFlorian:2003qf, Eskola:2001gt}, energy loss~\cite{Dai:2020rlu} and nuclear breakup~\cite{Capella:1996va,Capella:2006mb},
which produce effects similar to QGP. The study of particle production in small systems is thus crucial to disentangle CNM effects from the possible formation of QGP at temperature and density lower than it is generally expected~\cite{Armesto:2018ljh}.

One of the signatures of QGP formation is the enhancement of strange-hadron production as a result of the chemical equilibrium of strange quarks~\cite{PhysRevLett.48.1066}. This enhancement has been observed in a variety of collision systems~\cite{WA97:1999uwz, STAR:2007cqw,ALICE:2016fzo}. The $\phi$(1020) meson, hereafter referred to as $\phi$, is abundantly produced at the LHC because it is a light bound state of a strange quark-antiquark pair.
The $\phi$ meson production in ordinary matter is suppressed due to the Okubo--Zweig--Iizuka (OZI) rule compared to other vector mesons of similar mass \cite{Okubo:1963fa, Zweig:1964ruk, Iizuka:1966fk}. In the QGP medium, $\phi$ mesons can instead be produced through the coalescence of abundant strange quarks, bypassing the OZI rule. This increase makes the $\phi$ meson an ideal probe of QGP~\cite{Shor:1984ui}. Alternative mechanisms to QGP formation such as string fusion~\cite{Armesto:1994yg}, colour reconnection \cite{Christiansen:2015yqa} and rope hadronisation~\cite{Bierlich:2022ned} have also been proposed to describe strangeness enhancement.
Hence, studying $\phi$ meson production in proton-nucleus collisions at different centre-of-mass energies is essential to provide insights into this complex variety of scenarios.

This paper presents the first measurement of $\phi$ meson production in proton-neon ($\proton$Ne) fixed-target collisions at $\sqsnn=68.5$\gev by the \lhcb experiment. The differential production cross-section is measured as a function of the $\phi$ meson transverse momentum \pt, and rapidity in the centre-of-mass of the reaction $y^*$.

\section{The \lhcb detector and the SMOG system}

The \lhcb detector~\cite{LHCb-DP-2008-001,LHCb-DP-2014-002} is a single-arm forward spectrometer designed for studying particles containing $c$ or $b$ quarks, covering the pseudorapidity range $2 < \eta < 5$. The detector comprises a silicon-strip vertex locator (VELO), three tracking stations equipped with silicon-strip detectors and straw drift tubes, two ring-imaging Cherenkov (RICH) detectors for charged hadron identification, a calorimeter system with scintillating-pad and preshower detectors, electromagnetic and hadronic calorimeters, and a muon detector with alternating layers of iron and multiwire proportional chambers.

The System for Measuring the Overlap with Gas (SMOG)~\cite{LHCb-PAPER-2014-047} enables the injection of gases (He, Ar, Ne) at pressures of $\mathcal{O}(10^{-7})$ mbar into the beam pipe section within the VELO, allowing \lhcb to operate as a fixed-target experiment. This provides a unique opportunity to study proton-nucleus and nucleus-nucleus collisions with different gaseous targets using the LHC beams. In this configuration, the high-energy proton beam allows the \lhcb acceptance to extend into the negative rapidity hemisphere in the nucleon-nucleon centre-of-mass system of the reaction, covering the region from $y^* \sim -2.5$ to $y^* \sim 0$.

\section{Data sample and simulation}
\label{sec:data}

The $\proton$Ne sample was collected in 2017 simultaneously with the proton-proton ($\proton\proton$) collision data, where a beam of protons with an energy of 2.51\tev collided against neon nuclei at rest, resulting in a centre-of-mass energy of \sqsnn = 68.5\gev.

Simulated samples of $\phi \to \Kp\Km$ decays are generated with \pythia~\cite{Sjostrand:2007gs}, configured specifically for \lhcb~\cite{LHCb-PROC-2010-056}, with the colliding proton beam momentum set to match the momentum per nucleon of both the beam and the target in the centre-of-mass frame. 
The four-momentum of the $\phi$ decay products is then embedded into $\proton$Ne minimum bias events generated by the EPOS-LHC event generator~\cite{Pierog_2015}. The interaction of the generated particles with the detector, along with its response, are simulated using the \geant toolkit~\cite{Agostinelli:2002hh, Allison:2006ve}, as described in Ref.~\cite{Clemencic:2011zza}.
The simulated samples are reconstructed and analysed using the same software tools employed for data processing.

\section{Event selection}
\label{sec:sel}

The proton-gas events are selected by a two-stage trigger~\cite{LHCb-DP-2012-004}. The first stage is hardware-based and selects a constant rate of proton bunches moving towards the detector. These bunches do not cross any filled bunch moving in the opposite direction at the nominal \proton\proton interaction point.
The second stage is software-based and selects events with at least one track in the VELO. In addition, to suppress residual \proton\proton collisions, events with activity in the backward region are vetoed based on the number of tracks in VELO stations upstream of the interaction region.

In the offline selection, the $\phi$ candidates are formed from two opposite-charge kaon tracks forming a good-quality vertex close to the primary $\proton$Ne collision vertex (PV).
Each kaon track is required to have a minimum momentum of $2\gevc$ and a minimum transverse momentum of $400\mevc$, and the selected pairs should have an invariant mass between 990 and 1100\mevcc. Tight particle identification (PID) requirements are applied to select kaons.
The measurement is performed in the $\phi$ transverse momentum range $800<\pt<6500$\mevc and centre-of-mass rapidity range $-1.8<y^*<0$.
The $z$~coordinate of the PV is required to lie in the interval $z_{\rm{PV}} \in [-200,-100] \cup [100, 150] \mm$ to suppress the contamination from \proton\proton collisions that occur at $z_{\rm{PV}} \sim 0 \mm$.

\begin{figure}[!t]
  \begin{center}
    \includegraphics[width=\linewidth]{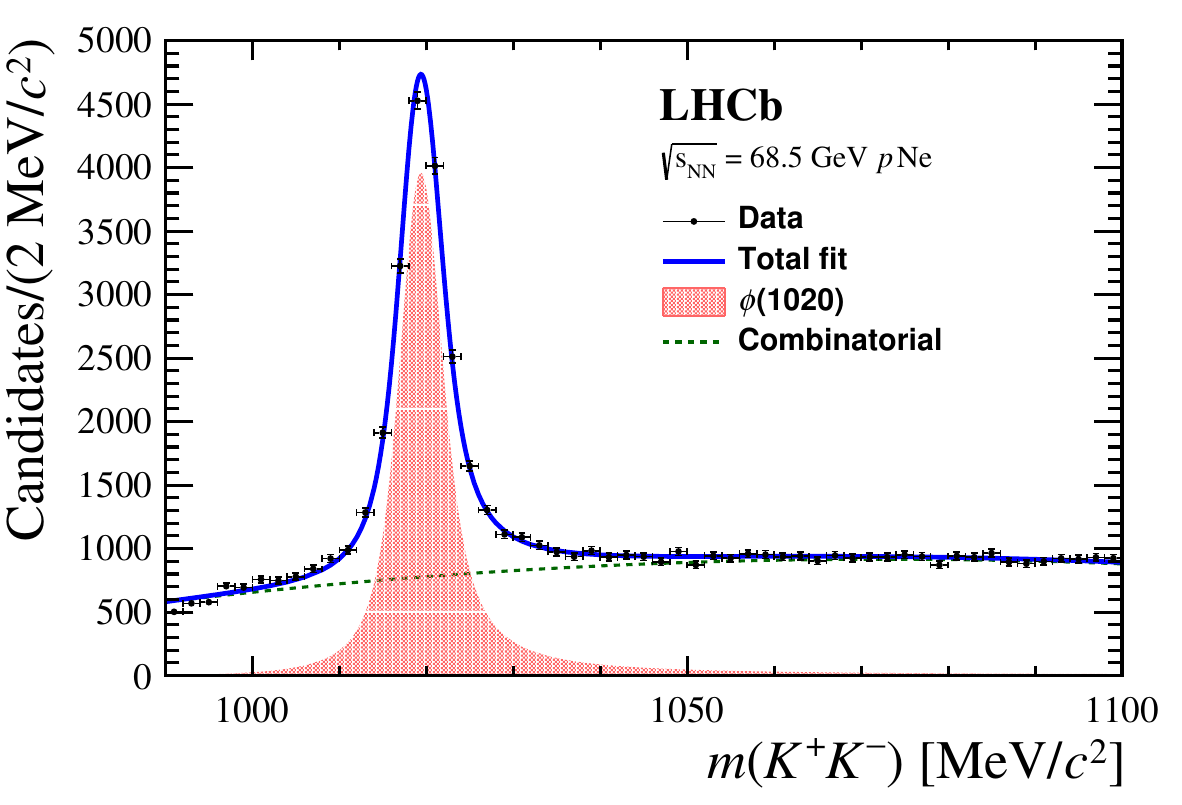}
  \end{center}
  \caption{Invariant-mass distribution of $\phi \rightarrow K^{+} K^{-}$ candidates after the selection. The fit curve (solid blue line) is composed of combinatorial background (dashed green line) and signal (red area).}
  \label{fig:MM}
\end{figure}

Figure~\ref{fig:MM} shows the $\Kp\Km$ invariant-mass distribution of the selected candidates.
The fit function is also shown and is composed of a background and a signal component. The background is described with a second-order polynomial function, where the coefficients are free to vary. The signal is described with a relativistic Breit--Wigner function,  
\begin{equation}
        f_{\rm{sig}}(m) = \frac{m\Gamma(m)}{(m^{2} - m_{\phi}^{2})^{2} + m_{\phi}^{2}\Gamma^2(m)},
        \label{eq:relBW}
\end{equation}
where
\begin{equation}
    \Gamma(m) = \Gamma_{\phi} \frac{m}{m_{\phi}} \left[ \frac{m^{2}/4-m_{K}^{2}}{m_{\phi}^{2}/4-m_{K}^{2}} \right]^{3/2}.
    \label{eq:gamma}
\end{equation}
In the fit, $m_K$, $m_\phi$ and $\Gamma_\phi$ parameters are fixed to the known values~\cite{PDG2024}. To account for detector resolution effects, Eq.~\ref{eq:relBW} is convolved with a Gaussian function whose width is free to vary in the fit.
A total of $16866\pm185$ signal events are obtained from the unbinned maximum-likelihood fit to the \Kp\Km invariant mass $(m)$. 

The double-differential cross-section per target nucleon is estimated by correcting the signal yield in each kinematic region, $N(\pt,y^*)$, obtained from independent fits, for the $\phi\to \Kp\Km$ decay branching fraction \mbox{$\BF(\phi \to \Kp\Km)=49.1\pm0.5\%$}~\cite{PDG2024}, the integrated luminosity of the sample $\mathcal{L}$, and the total efficiency $\epsilon_{\rm tot}(\pt,y^*)$ according to the formula
\begin{equation}
    \frac{\deriv^2\sigma}{\deriv\pt \deriv y^*} = \frac{1}{\Delta \pt \Delta y^*}\frac{N(\pt,y^*)}{\epsilon_{\rm tot}(\pt,y^*) \cdot \mathcal{L} \cdot\BF(\phi \to \Kp\Km)\cdot A} \, ,
\end{equation}
where $\Delta \pt$ and $\Delta y$ are the kinematic region widths and $A=20.18$ is the atomic weight of neon~\cite{atomic_weight}.
The total efficiency includes the geometrical acceptance of the detector and the efficiency of trigger, event reconstruction and offline selection.

The PV reconstruction efficiency is measured using a $\phi \to \Kp\Km$ data sample reconstructed without vertex requirements~\cite{LHCb-PAPER-2022-014}. The PID efficiency is estimated from a $\phi \to \Kp\Km$ data sample where one of the two kaons (the \textit{tag}) is selected with a tight PID requirement, while the other one (the \textit{probe}) is selected without any PID condition.
The efficiency for \Kp and \Km is defined as the fraction of $\phi$ candidates in which the probe track passes the PID condition, which is estimated from an invariant-mass fit. The remaining efficiencies are computed from simulated samples, where events are assigned weights to ensure that the distribution of VELO cluster multiplicity matches that observed in data. Data corrections from \proton\proton calibration samples are also applied when estimating the track reconstruction efficiency. The integrated luminosity is determined to be $21.7\pm1.4$\invnb from the elastic proton scattering on neon atomic electrons~\cite{LHCb-PAPER-2018-031}.

\section{Systematic uncertainties}
\label{sec:sys}
Several sources of systematic uncertainty are considered, including those associated with the signal and background fit models, as well as with the evaluation of the efficiencies. 

A systematic uncertainty is evaluated by computing the relative difference between the signal yield obtained with the baseline fit model and an alternative model based on an exponential function for the background and a Breit--Wigner function for the signal that includes the Blatt--Weisskopf barrier factors. These factors modify the mass-dependent signal width (Eq.~\ref{eq:gamma}) into 
\begin{equation}
    \Gamma(m) = \Gamma_{\phi} \frac{m}{m_{\phi}} \left[ \frac{m^{2}/4-m_{K}^{2}}{m_{\phi}^{2}/4-m_{K}^{2}} \right]^{3/2}
    \frac{1+r_0^2\left(
    m_{\phi}^2/4-m^2_K
    \right)}{1+r_0^2\left(m^2/4-m^2_K\right)},
\end{equation}
where the radius of the $\phi$ meson is fixed to $r_0=5.1~\gev^{-1}$~\cite{PDG2024}. 

The uncertainty on the geometrical acceptance is due to the limited size of the simulated sample.
An alternative set of weights to correct the multiplicity in the simulation is obtained using the number of tracks in the event instead of the number of VELO clusters. The difference between the efficiencies computed with this set of weights and the nominal efficiencies is assigned as a systematic uncertainty.
The uncertainty on the PV reconstruction and PID efficiencies arise from the size of the individual data sample used for each estimate. The uncertainty on the reconstruction and selection efficiency depends on both the size of the simulated sample and the uncertainty on the tracking corrections obtained from data~\cite{LHCb-DP-2013-002}. An additional source of systematic uncertainty arises from the contamination of residual \proton\proton collisions, which is evaluated from the comparison of data samples of pure $\proton$Ne and pure \proton\proton collisions~\cite{LHCb-PAPER-2022-014}.  Finally, the uncertainty on the $\phi \to \Kp\Km$ branching ratio and on the estimate of the integrated luminosity are also considered.

The sources and the corresponding uncertainty values are summarised in Table~\ref{tab:sysrecap}, separated between those correlated and uncorrelated across the kinematic regions.

\begin{table}[!tb]
  \centering
  \caption{Summary of the relative contributions to the systematic and statistical uncertainties for this measurement. Correlated systematic uncertainties affect all kinematic regions by the same amount, while ranges denote the minimum and maximum values among the \pt and $y^*$ regions.
}

\begin{tabular}{l c}
\hline
Systematic uncertainties & \\
\hline
\multicolumn{2}{c}{Uncorrelated among kinematic regions}\\
\hline
Signal determination & ($<$0.1 \text{--} 5.9)\% \\
Geometrical acceptance & ($<$0.1 \text{--} 1.8)\%\\
Multiplicity corrections & (0.2 \text{--} 3.6)\% \\
PID efficiency & (2.3 \text{--} 4.4)\%\\
\hline
\multicolumn{2}{c}{Correlated among kinematic regions}\\
\hline
PV reconstruction & 1.2\%\\
Reconstruction and selection efficiency & 3.2\%\\
\proton\proton contamination & 2.0\%\\
$\BF(\phi \to \Kp\Km)$ & 1.0\%\\
Luminosity & 6.5\% \\
\hline
Statistical uncertainty & (3.7 \text{--} 6.7)\%\\
\hline
\end{tabular}

\label{tab:sysrecap}
\end{table}

\section{Results}
\label{sec:results}

\begin{figure}[tb!]
	\centering 
	\includegraphics[width=0.95\textwidth]{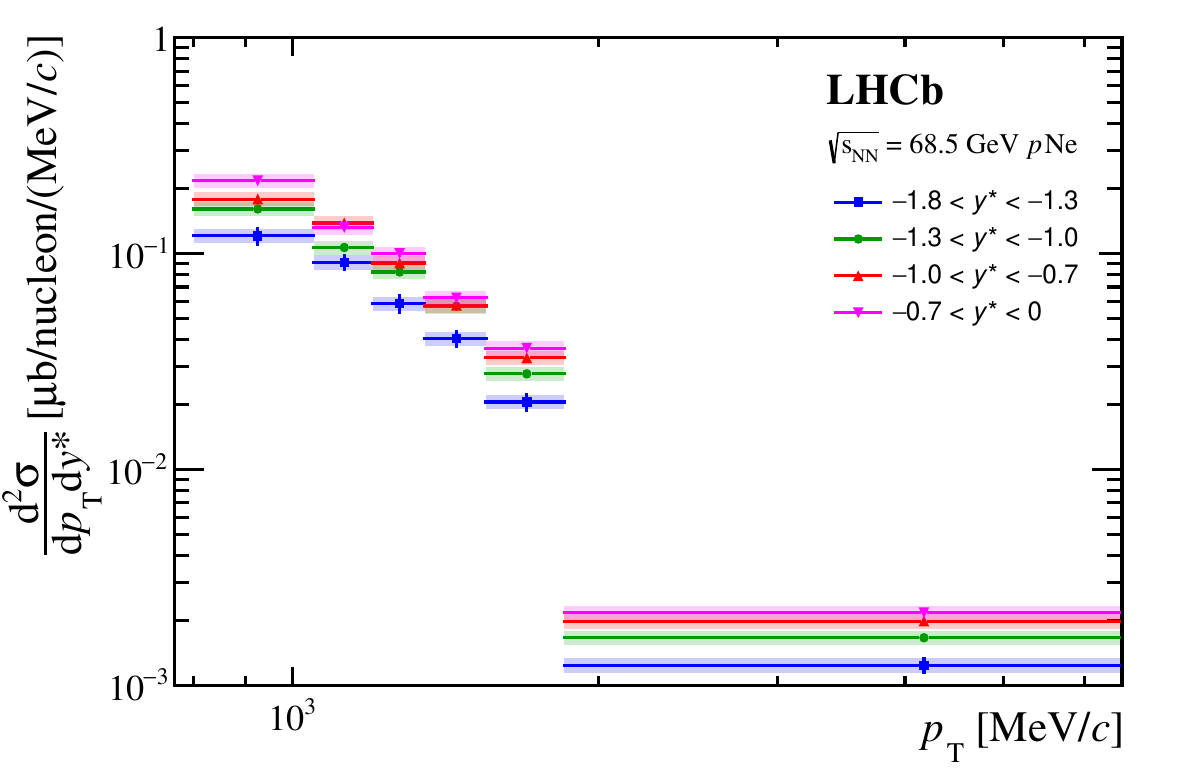}
	\caption{Double-differential $\phi$-meson production cross-section per nucleon. Statistical and uncorrelated uncertainties are indicated by the bars, while the filled area represents the correlated uncertainty.}
	\label{fig:xsec}
\end{figure}

\begin{figure}[tb!]
\begin{center}
    \includegraphics[width=0.89\linewidth]{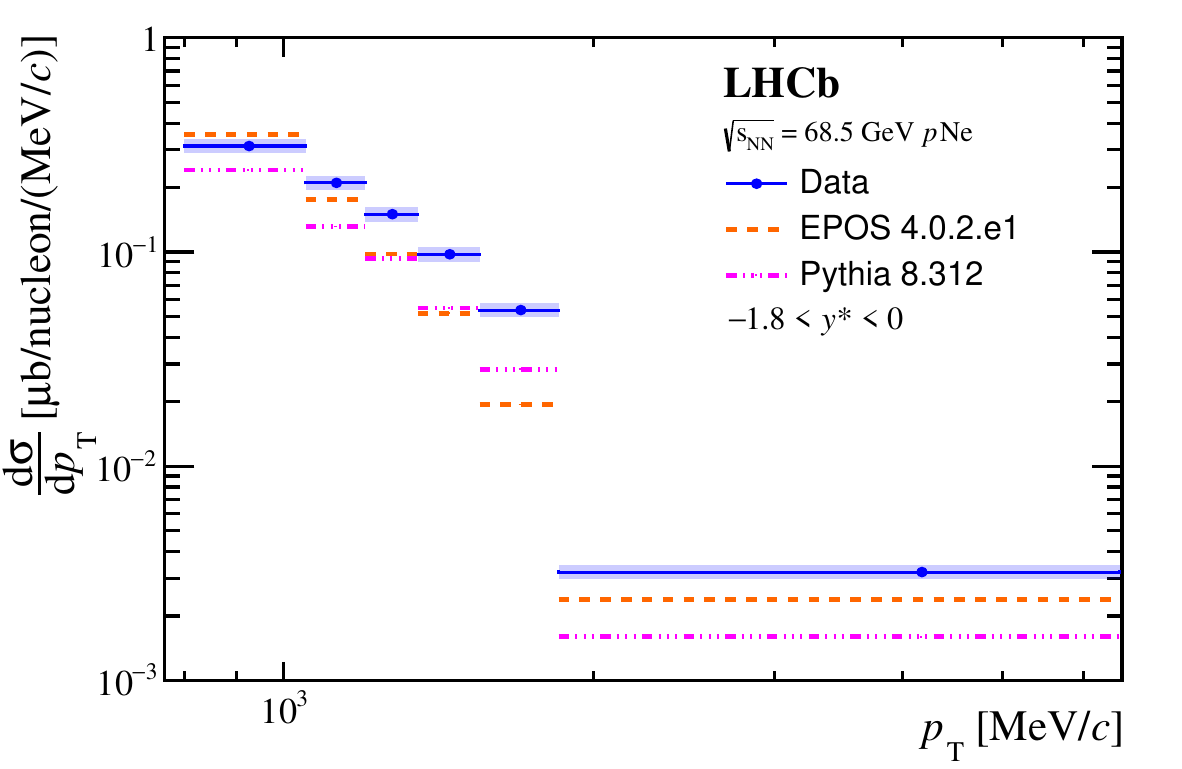}
    \includegraphics[width=0.7\linewidth]{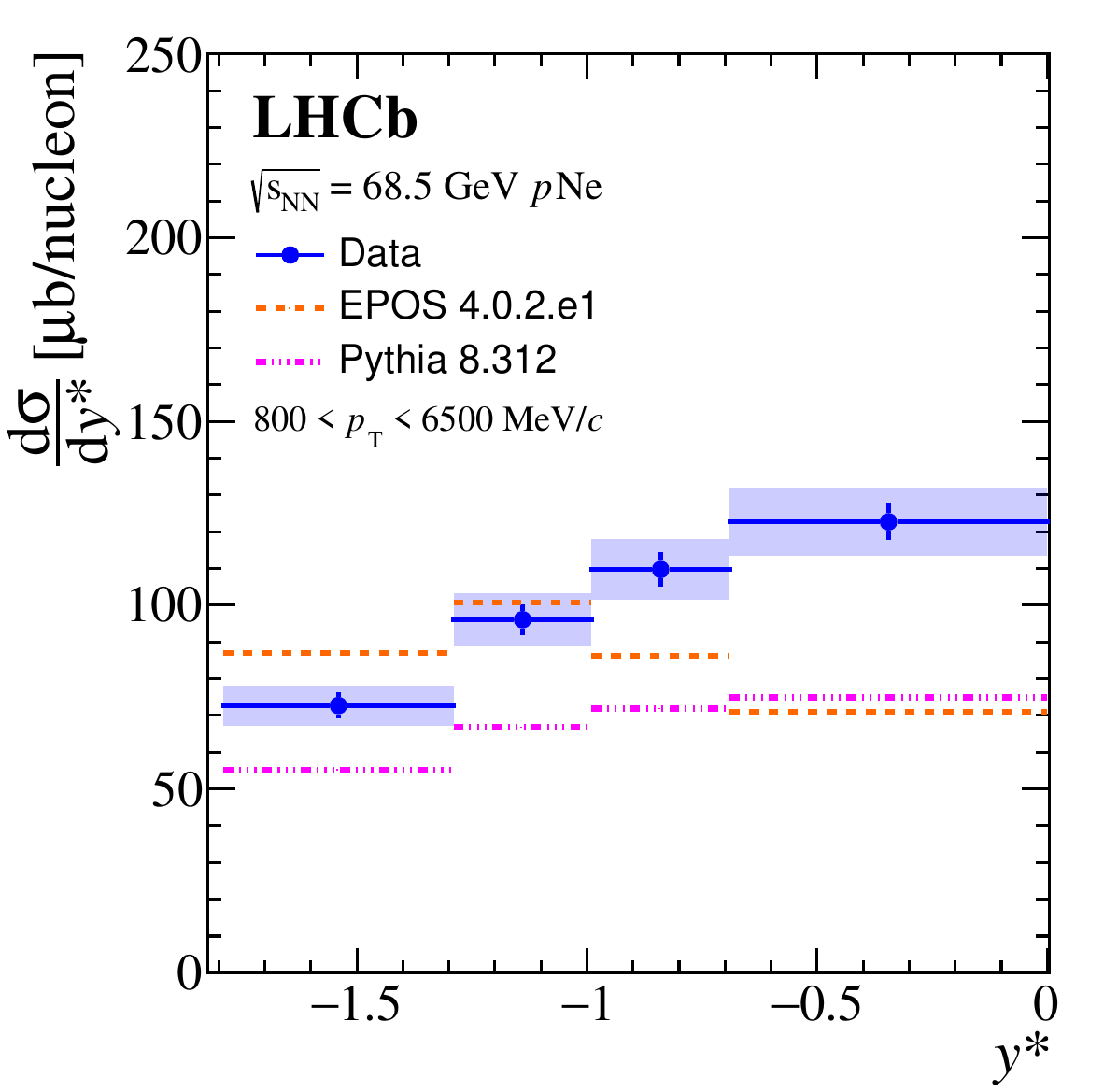}
    \vspace*{-0.5cm}
  \end{center}
\caption{Single-differential cross-sections per nucleon (blue) as a function of (top) \pt in $-1.8<y^*<0$, and (bottom) $y^*$ in $800<\textit{p}_{\text{T}}<6500$ MeV/\textit{c}. The measurement are compared to EPOS4 (orange) and Pythia 8.312 (magenta) predictions. On the measurement, statistical and uncorrelated uncertainties are indicated by the bars, while the filled area represents the correlated uncertainty.}
\label{fig:xsecsin} 
\end{figure}

The measured $\phi$ production cross-section in $\proton$Ne fixed-target collisions at \sqsnn = 68.5\gev within $-1.8<y^*<0$ and $800<\pt<6500$\mevc is found to be
\begin{equation*}
    \sigma = 182.7\pm2.7~\text{(stat.)}\pm14.1~\text{(syst.)}\text{\mub/nucleon}.
\end{equation*}
The double-differential cross-section is shown in Fig.~\ref{fig:xsec}, in different rapidity ranges as a function of \pt. 
The cross-section shows a decreasing trend with increasing momentum and large negative rapidity, as observed for charmed mesons with the same energy and collision system~\cite{LHCb-PAPER-2022-014}.
By integrating the results, the single-differential cross-sections as a function of \pt and $y^*$ are obtained and shown in Fig.~\ref{fig:xsecsin} together with the predictions from Pythia 8.312~\cite{Bierlich:2022pfr} and from a preliminary EPOS4 release~\cite{Werner:2023zvo, Werner:2023fne, Werner:2023mod, Werner:2023jps}. 
This version of Pythia includes the Angantyr model~\cite{Bierlich:2018xfw}, which can simulate 
$\proton\proton$, $\proton$A, and AA collisions, and offers the flexibility to modify the initial-state geometry model to accommodate light nuclei such as neon. In comparison, the preliminary release of EPOS4 provides pregenerated tables specifically for $\proton$Ne collisions at \sqsnn = 68.5\gev. The strong \pt dependence is reproduced by the models, although they generally underestimate the data in almost all intervals.

The numerical values of the cross-section are reported in Tables~\ref{tab:xsec} and \ref{tab:xsecsin}. The average transverse momentum in each centre-of-mass rapidity region is provided in Table~\ref{tab:ptavg}.

\section{Conclusions}\label{sec:conclu}

The first measurement of the production cross-section of the $\phi(1020)$ meson in $\proton$Ne collisions at $\sqsnn=68.5$\gev by the \lhcb experiment is presented. 
The cross-section in the range of rapidity $-1.8<y^*<0$ and transverse momentum $800<\pt<6500$\mevc is found to be
\begin{equation*}
    \sigma = 182.7\pm2.7~\rm{(stat.)}\pm14.1~\rm{(syst.)}\rm{\mub/nucleon},
\end{equation*}
where the first uncertainty is statistical, and the second is systematic. 
The differential cross-section is compared to predictions based on Pythia 8.312 and preliminary EPOS4 generators, which generally underestimate the experimental points. 
This result provides useful information to tune phenomenological models on strangeness production and will constitute a key reference for QGP searches in heavier colliding systems.

\begin{table}[tb!]
  \centering
  \caption{Double-differential cross-section per nucleon as a function of $(\pt, y^{*})$ in units of \mub/nucleon/(\mevc). The first uncertainty includes the statistical and uncorrelated contributions, while the second represents the correlated uncertainty.}

\begin{tabular}{c| c c}
Range & \multicolumn{2}{c}{$y^{*}$} \\
\hline
$\pt$ [\mevc] & $[-1.8, -1.3]$ & $[-1.3, -1.0]$ \\
\hline
[\phantom{0}800, 1050] & 0.121$\pm$0.010$\pm$0.009 & 0.161$\pm$0.012$\pm$0.012 \\

[1050, 1200] & 0.091$\pm$0.007$\pm$0.007 & 0.107$\pm$0.008$\pm$0.008 \\

[1200, 1350] & 0.059$\pm$0.005$\pm$0.004 & 0.082$\pm$0.006$\pm$0.006 \\

[1350, 1550] & 0.040$\pm$0.003$\pm$0.003 & 0.057$\pm$0.004$\pm$0.004 \\

[1550, 1850] & 0.021$\pm$0.002$\pm$0.002 & 0.028$\pm$0.002$\pm$0.002 \\

[1850, 6500] & 0.0012$\pm$0.0001$\pm$0.0001 & 0.0017$\pm$0.0001$\pm$0.0001 \\
\hline
\end{tabular}

\vspace{0.6cm}

\begin{tabular}{c| c c}
Range & \multicolumn{2}{c}{$y^{*}$} \\
\hline
$\pt$ [\mevc] & $[-1.0, -0.7]$ & $[-0.7, 0.0]$ \\
\hline
[\phantom{0}800, 1050] & 0.179$\pm$0.013$\pm$0.014 & 0.218$\pm$0.013$\pm$0.014 \\

[1050, 1200] & 0.139$\pm$0.011$\pm$0.011 & 0.132$\pm$0.008$\pm$0.010 \\

[1200, 1350] & 0.090$\pm$0.006$\pm$0.007 & 0.100$\pm$0.006$\pm$0.008 \\

[1350, 1550] & 0.058$\pm$0.005$\pm$0.004 & 0.062$\pm$0.004$\pm$0.005 \\

[1550, 1850] & 0.033$\pm$0.002$\pm$0.003 & 0.036$\pm$0.003$\pm$0.003 \\

[1850, 6500] & 0.0020$\pm$0.0002$\pm$0.0002 & 0.0022$\pm$0.0001$\pm$0.0002 \\
\hline
\end{tabular}
\label{tab:xsec}
\end{table}

\begin{table}[tb]
  \centering
  \caption{Single-differential cross-section as a function of $\pt$ or $y^*$. The first uncertainty includes the statistical and uncorrelated contributions, while the second represents the correlated uncertainty.}

\begin{tabular}{c| c}

Range & Values\\
\hline
\pt [\mevc] & d$\sigma$/d\pt [\mub/nucleon/(\mevc)]\\
\hline
[\phantom{0}800, 1050] & 0.313$\pm$0.013$\pm$0.024 \\

[1050, 1200] & 0.210$\pm$0.008$\pm$0.016 \\

[1200, 1350] & 0.150$\pm$0.006$\pm$0.011 \\

[1350, 1550] & 0.098$\pm$0.004$\pm$0.007 \\

[1550, 1850] & 0.054$\pm$0.002$\pm$0.004 \\

[1850, 6500] & 0.0032$\pm$0.0001$\pm$0.0002 \\
\hline
\hline
$y^*$ & d$\sigma$/d$y^*$ [\mub/nucleon]\\
\hline
$[-1.8, -1.3]$ & 72.7$\pm$3.0$\pm$5.5 \\

$[-1.3, -1.0]$ & 96.1$\pm$3.6$\pm$7.3 \\

$[-1.0, -0.7]$ & 109.7$\pm$4.1$\pm$8.3 \\

$[-0.7, \phantom{-}0.0]$ & 122.7$\pm$4.3$\pm$9.3  \\
\hline
\end{tabular}

\label{tab:xsecsin}
\end{table}

\begin{table}[tb]
  \centering
  \caption{Average \pt in \mevc units of the $\phi$ meson in each kinematic region and in the full rapidity range (last column).}

\begin{tabular}{c| c c c c | c}

Range & & $y^{*}$& & & $y^{*}$ \\
\hline
\pt [\mevc] & $[-1.8, -1.3]$ & $[-1.3, -1.0]$& $[-1.0, -0.7]$& $[-0.7, 0.0]$ & $[-1.8, 0.0]$\\
\hline
{[\phantom{0}800, 1050]} & \phantom{0}961 &  \phantom{0}965 &  \phantom{0}951 & \phantom{0}964 & \phantom{0}960\\
{[1050, 1200]} & 1125 & 1125 & 1126 & 1127 & 1126\\
{[1200, 1350]} & 1274 & 1274 & 1272 & 1274 & 1274\\
{[1350, 1550]} & 1437 & 1442& 1447 & 1448 & 1444\\
{[1550, 1850]} & 1678 & 1682 & 1676 & 1688 & 1681\\
{[1850, 6500]} & 2255 & 2284 & 2259 & 2295 & 2273\\
\hline
\end{tabular}
\label{tab:ptavg}
\end{table}

% Do not include this in any draft (just for information in the template)
\section*{Acknowledgements}
%
% These Acknowledgements valid from 3-May-2019
%
\noindent 
We thank Christian Bierlich for his assistance with the Pythia prediction, and Klaus Werner and Mahbobeh Jafarpour for providing an EPOS release to generate $p$Ne fixed-target collisions.
We express our gratitude to our colleagues in the CERN
accelerator departments for the excellent performance of the LHC. We
thank the technical and administrative staff at the LHCb
institutes.
We express our gratitude to our colleagues in the CERN
accelerator departments for the excellent performance of the LHC. We
thank the technical and administrative staff at the LHCb
institutes.
We acknowledge support from CERN and from the national agencies:
CAPES, CNPq, FAPERJ and FINEP (Brazil); 
MOST and NSFC (China); 
CNRS/IN2P3 (France); 
BMBF, DFG and MPG (Germany); 
INFN (Italy); 
NWO (Netherlands); 
MNiSW and NCN (Poland); 
MCID/IFA (Romania); 
%MSHE (Russia); 
MICIU and AEI (Spain);
SNSF and SER (Switzerland); 
NASU (Ukraine); 
STFC (United Kingdom); 
DOE NP and NSF (USA).
We acknowledge the computing resources that are provided by CERN, IN2P3
(France), KIT and DESY (Germany), INFN (Italy), SURF (Netherlands),
PIC (Spain), GridPP (United Kingdom), 
%RRCKI and Yandex LLC (Russia), 
CSCS (Switzerland), IFIN-HH (Romania), CBPF (Brazil),
and Polish WLCG (Poland).
We are indebted to the communities behind the multiple open-source
software packages on which we depend.
Individual groups or members have received support from
ARC and ARDC (Australia);
Key Research Program of Frontier Sciences of CAS, CAS PIFI, CAS CCEPP, 
Fundamental Research Funds for the Central Universities, 
and Sci. \& Tech. Program of Guangzhou (China);
Minciencias (Colombia);
EPLANET, Marie Sk\l{}odowska-Curie Actions, ERC and NextGenerationEU (European Union);
A*MIDEX, ANR, IPhU and Labex P2IO, and R\'{e}gion Auvergne-Rh\^{o}ne-Alpes (France);
%RFBR, RSF and Yandex LLC (Russia);
AvH Foundation (Germany);
ICSC (Italy); 
%GVA, XuntaGal, GENCAT, Inditex, InTalent and Prog.~Atracci\'on Talento, CM (Spain);
Severo Ochoa and Mar\'ia de Maeztu Units of Excellence, GVA, XuntaGal, GENCAT, InTalent-Inditex and Prog. ~Atracci\'on Talento CM (Spain);
SRC (Sweden);
the Leverhulme Trust, the Royal Society
 and UKRI (United Kingdom).

\addcontentsline{toc}{section}{References}
%\setboolean{inbibliography}{true}
\bibliographystyle{LHCb}
\bibliography{main,standard,LHCb-PAPER,LHCb-CONF,LHCb-DP,LHCb-TDR}

\ifx\mcitethebibliography\mciteundefinedmacro
\PackageError{LHCb.bst}{mciteplus.sty has not been loaded}
{This bibstyle requires the use of the mciteplus package.}\fi
\providecommand{\href}[2]{#2}
\begin{mcitethebibliography}{10}
\mciteSetBstSublistMode{n}
\mciteSetBstMaxWidthForm{subitem}{\alph{mcitesubitemcount})}
\mciteSetBstSublistLabelBeginEnd{\mcitemaxwidthsubitemform\space}
{\relax}{\relax}

\bibitem{Shuryak:1980tp}
E.~V. Shuryak, \ifthenelse{\boolean{articletitles}}{\emph{{Quantum chromodynamics and the theory of superdense matter}}, }{}\href{https://doi.org/10.1016/0370-1573(80)90105-2}{Phys.\ Rept.\  \textbf{61} (1980) 71}\relax
\mciteBstWouldAddEndPuncttrue
\mciteSetBstMidEndSepPunct{\mcitedefaultmidpunct}
{\mcitedefaultendpunct}{\mcitedefaultseppunct}\relax
\EndOfBibitem
\bibitem{STAR:2005gfr}
STAR collaboration, J.~Adams {\em et~al.}, \ifthenelse{\boolean{articletitles}}{\emph{{Experimental and theoretical challenges in the search for the quark-gluon plasma: The STAR collaboration's critical assessment of the evidence from RHIC collisions}}, }{}\href{https://doi.org/10.1016/j.nuclphysa.2005.03.085}{Nucl.\ Phys.\  \textbf{A757} (2005) 102}, \href{http://arxiv.org/abs/nucl-ex/0501009}{{\normalfont\ttfamily arXiv:nucl-ex/0501009}}\relax
\mciteBstWouldAddEndPuncttrue
\mciteSetBstMidEndSepPunct{\mcitedefaultmidpunct}
{\mcitedefaultendpunct}{\mcitedefaultseppunct}\relax
\EndOfBibitem
\bibitem{PHENIX:2004vcz}
PHENIX collaboration, K.~Adcox {\em et~al.}, \ifthenelse{\boolean{articletitles}}{\emph{{Formation of dense partonic matter in relativistic nucleus-nucleus collisions at RHIC: Experimental evaluation by the PHENIX collaboration}}, }{}\href{https://doi.org/10.1016/j.nuclphysa.2005.03.086}{Nucl.\ Phys.\  \textbf{A757} (2005) 184}, \href{http://arxiv.org/abs/nucl-ex/0410003}{{\normalfont\ttfamily arXiv:nucl-ex/0410003}}\relax
\mciteBstWouldAddEndPuncttrue
\mciteSetBstMidEndSepPunct{\mcitedefaultmidpunct}
{\mcitedefaultendpunct}{\mcitedefaultseppunct}\relax
\EndOfBibitem
\bibitem{ALICE:2022wpn}
ALICE collaboration, S.~Acharya {\em et~al.}, \ifthenelse{\boolean{articletitles}}{\emph{{The ALICE experiment: a journey through QCD}}, }{}\href{https://doi.org/10.1140/epjc/s10052-024-12935-y}{Eur.\ Phys.\ J.\  \textbf{C84} (2024) 813}, \href{http://arxiv.org/abs/2211.04384}{{\normalfont\ttfamily arXiv:2211.04384}}\relax
\mciteBstWouldAddEndPuncttrue
\mciteSetBstMidEndSepPunct{\mcitedefaultmidpunct}
{\mcitedefaultendpunct}{\mcitedefaultseppunct}\relax
\EndOfBibitem
\bibitem{PHENIX:2018lia}
PHENIX collaboration, C.~Aidala {\em et~al.}, \ifthenelse{\boolean{articletitles}}{\emph{{Creation of quark-gluon plasma droplets with three distinct geometries}}, }{}\href{https://doi.org/10.1038/s41567-018-0360-0}{Nature Phys.\  \textbf{15} (2019) 214}, \href{http://arxiv.org/abs/1805.02973}{{\normalfont\ttfamily arXiv:1805.02973}}\relax
\mciteBstWouldAddEndPuncttrue
\mciteSetBstMidEndSepPunct{\mcitedefaultmidpunct}
{\mcitedefaultendpunct}{\mcitedefaultseppunct}\relax
\EndOfBibitem
\bibitem{deFlorian:2003qf}
D.~de~Florian and R.~Sassot, \ifthenelse{\boolean{articletitles}}{\emph{{Nuclear parton distributions at next to leading order}}, }{}\href{https://doi.org/10.1103/PhysRevD.69.074028}{Phys.\ Rev.\  \textbf{D69} (2004) 074028}, \href{http://arxiv.org/abs/hep-ph/0311227}{{\normalfont\ttfamily arXiv:hep-ph/0311227}}\relax
\mciteBstWouldAddEndPuncttrue
\mciteSetBstMidEndSepPunct{\mcitedefaultmidpunct}
{\mcitedefaultendpunct}{\mcitedefaultseppunct}\relax
\EndOfBibitem
\bibitem{Eskola:2001gt}
K.~J. Eskola, V.~J. Kolhinen, and R.~Vogt, \ifthenelse{\boolean{articletitles}}{\emph{{Obtaining the nuclear gluon distribution from heavy quark decays to lepton pairs in pA collisions}}, }{}\href{https://doi.org/10.1016/S0375-9474(01)01221-0}{Nucl.\ Phys.\  \textbf{A696} (2001) 729}, \href{http://arxiv.org/abs/hep-ph/0104124}{{\normalfont\ttfamily arXiv:hep-ph/0104124}}\relax
\mciteBstWouldAddEndPuncttrue
\mciteSetBstMidEndSepPunct{\mcitedefaultmidpunct}
{\mcitedefaultendpunct}{\mcitedefaultseppunct}\relax
\EndOfBibitem
\bibitem{Dai:2020rlu}
T.~Dai, J.-F. Paquet, D.~Teaney, and S.~A. Bass, \ifthenelse{\boolean{articletitles}}{\emph{{Parton energy loss in a hard-soft factorized approach}}, }{}\href{https://doi.org/10.1103/PhysRevC.105.034905}{Phys.\ Rev.\  \textbf{C105} (2022) 034905}, \href{http://arxiv.org/abs/2012.03441}{{\normalfont\ttfamily arXiv:2012.03441}}\relax
\mciteBstWouldAddEndPuncttrue
\mciteSetBstMidEndSepPunct{\mcitedefaultmidpunct}
{\mcitedefaultendpunct}{\mcitedefaultseppunct}\relax
\EndOfBibitem
\bibitem{Capella:1996va}
A.~Capella, A.~Kaidalov, A.~Kouider~Akil, and C.~Gerschel, \ifthenelse{\boolean{articletitles}}{\emph{{\jpsi and $\psi^\prime$ suppression in heavy ion collisions}}, }{}\href{https://doi.org/10.1016/S0370-2693(96)01650-4}{Phys.\ Lett.\  \textbf{B393} (1997) 431}, \href{http://arxiv.org/abs/hep-ph/9607265}{{\normalfont\ttfamily arXiv:hep-ph/9607265}}\relax
\mciteBstWouldAddEndPuncttrue
\mciteSetBstMidEndSepPunct{\mcitedefaultmidpunct}
{\mcitedefaultendpunct}{\mcitedefaultseppunct}\relax
\EndOfBibitem
\bibitem{Capella:2006mb}
A.~Capella, E.~G. Ferreiro, and E.~G. Ferreiro, \ifthenelse{\boolean{articletitles}}{\emph{{\jpsi suppression and the decrease of nuclear absorption with increasing energy}}, }{}\href{https://doi.org/10.1103/PhysRevC.76.064906}{Phys.\ Rev.\  \textbf{C76} (2007) 064906}, \href{http://arxiv.org/abs/hep-ph/0610313}{{\normalfont\ttfamily arXiv:hep-ph/0610313}}\relax
\mciteBstWouldAddEndPuncttrue
\mciteSetBstMidEndSepPunct{\mcitedefaultmidpunct}
{\mcitedefaultendpunct}{\mcitedefaultseppunct}\relax
\EndOfBibitem
\bibitem{Armesto:2018ljh}
N.~Armesto, \ifthenelse{\boolean{articletitles}}{\emph{{Small collision systems: theory overview on cold nuclear matter effects}}, }{}\href{https://doi.org/10.1051/epjconf/201817111001}{EPJ Web Conf.\  \textbf{171} (2018) 11001}\relax
\mciteBstWouldAddEndPuncttrue
\mciteSetBstMidEndSepPunct{\mcitedefaultmidpunct}
{\mcitedefaultendpunct}{\mcitedefaultseppunct}\relax
\EndOfBibitem
\bibitem{PhysRevLett.48.1066}
J.~Rafelski and B.~M\"uller, \ifthenelse{\boolean{articletitles}}{\emph{Strangeness production in the quark-gluon plasma}, }{}\href{https://doi.org/10.1103/PhysRevLett.48.1066}{Phys.\ Rev.\ Lett.\  \textbf{48} (1982) 1066}\relax
\mciteBstWouldAddEndPuncttrue
\mciteSetBstMidEndSepPunct{\mcitedefaultmidpunct}
{\mcitedefaultendpunct}{\mcitedefaultseppunct}\relax
\EndOfBibitem
\bibitem{WA97:1999uwz}
WA97 collaboration, E.~Andersen {\em et~al.}, \ifthenelse{\boolean{articletitles}}{\emph{{Strangeness enhancement at mid-rapidity in Pb-Pb collisions at 158 A\gevc}}, }{}\href{https://doi.org/10.1016/S0370-2693(99)00140-9}{Phys.\ Lett.\  \textbf{B449} (1999) 401}\relax
\mciteBstWouldAddEndPuncttrue
\mciteSetBstMidEndSepPunct{\mcitedefaultmidpunct}
{\mcitedefaultendpunct}{\mcitedefaultseppunct}\relax
\EndOfBibitem
\bibitem{STAR:2007cqw}
STAR collaboration, B.~I. Abelev {\em et~al.}, \ifthenelse{\boolean{articletitles}}{\emph{{Enhanced strange baryon production in Au+Au collisions compared to p+p at \sqsnn = 200\gev}}, }{}\href{https://doi.org/10.1103/PhysRevC.77.044908}{Phys.\ Rev.\  \textbf{C77} (2008) 044908}, \href{http://arxiv.org/abs/0705.2511}{{\normalfont\ttfamily arXiv:0705.2511}}\relax
\mciteBstWouldAddEndPuncttrue
\mciteSetBstMidEndSepPunct{\mcitedefaultmidpunct}
{\mcitedefaultendpunct}{\mcitedefaultseppunct}\relax
\EndOfBibitem
\bibitem{ALICE:2016fzo}
ALICE collaboration, J.~Adam {\em et~al.}, \ifthenelse{\boolean{articletitles}}{\emph{{Enhanced production of multi-strange hadrons in high-multiplicity proton-proton collisions}}, }{}\href{https://doi.org/10.1038/nphys4111}{Nature Phys.\  \textbf{13} (2017) 535}, \href{http://arxiv.org/abs/1606.07424}{{\normalfont\ttfamily arXiv:1606.07424}}\relax
\mciteBstWouldAddEndPuncttrue
\mciteSetBstMidEndSepPunct{\mcitedefaultmidpunct}
{\mcitedefaultendpunct}{\mcitedefaultseppunct}\relax
\EndOfBibitem
\bibitem{Okubo:1963fa}
S.~Okubo, \ifthenelse{\boolean{articletitles}}{\emph{{$\phi$ meson and unitary symmetry model}}, }{}\href{https://doi.org/10.1016/S0375-9601(63)92548-9}{Phys.\ Lett.\  \textbf{5} (1963) 165}\relax
\mciteBstWouldAddEndPuncttrue
\mciteSetBstMidEndSepPunct{\mcitedefaultmidpunct}
{\mcitedefaultendpunct}{\mcitedefaultseppunct}\relax
\EndOfBibitem
\bibitem{Zweig:1964ruk}
G.~Zweig, \ifthenelse{\boolean{articletitles}}{\emph{{An $SU_3$ model for strong interaction symmetry and its breaking. Version 1}}, }{}doi:~\href{https://doi.org/10.17181/CERN-TH-401}{10.17181/CERN-TH-401}\relax
\mciteBstWouldAddEndPuncttrue
\mciteSetBstMidEndSepPunct{\mcitedefaultmidpunct}
{\mcitedefaultendpunct}{\mcitedefaultseppunct}\relax
\EndOfBibitem
\bibitem{Iizuka:1966fk}
J.~Iizuka, \ifthenelse{\boolean{articletitles}}{\emph{{Systematics and phenomenology of meson family}}, }{}\href{https://doi.org/10.1143/PTPS.37.21}{Prog.\ Theor.\ Phys.\ Suppl.\  \textbf{37} (1966) 21}\relax
\mciteBstWouldAddEndPuncttrue
\mciteSetBstMidEndSepPunct{\mcitedefaultmidpunct}
{\mcitedefaultendpunct}{\mcitedefaultseppunct}\relax
\EndOfBibitem
\bibitem{Shor:1984ui}
A.~Shor, \ifthenelse{\boolean{articletitles}}{\emph{$\ensuremath{\phi}$-meson production as a probe of the quark-gluon plasma}, }{}\href{https://doi.org/10.1103/PhysRevLett.54.1122}{Phys.\ Rev.\ Lett.\  \textbf{54} (1985) 1122}\relax
\mciteBstWouldAddEndPuncttrue
\mciteSetBstMidEndSepPunct{\mcitedefaultmidpunct}
{\mcitedefaultendpunct}{\mcitedefaultseppunct}\relax
\EndOfBibitem
\bibitem{Armesto:1994yg}
N.~Armesto, M.~A. Braun, E.~G. Ferreiro, and C.~Pajares, \ifthenelse{\boolean{articletitles}}{\emph{{Strangeness enhancement and string fusion in nucleus-nucleus collisions}}, }{}\href{https://doi.org/10.1016/0370-2693(94)01511-A}{Phys.\ Lett.\  \textbf{B344} (1995) 301}\relax
\mciteBstWouldAddEndPuncttrue
\mciteSetBstMidEndSepPunct{\mcitedefaultmidpunct}
{\mcitedefaultendpunct}{\mcitedefaultseppunct}\relax
\EndOfBibitem
\bibitem{Christiansen:2015yqa}
J.~R. Christiansen and P.~Z. Skands, \ifthenelse{\boolean{articletitles}}{\emph{{String formation beyond leading colour}}, }{}\href{https://doi.org/10.1007/JHEP08(2015)003}{JHEP \textbf{08} (2015) 003}, \href{http://arxiv.org/abs/1505.01681}{{\normalfont\ttfamily arXiv:1505.01681}}\relax
\mciteBstWouldAddEndPuncttrue
\mciteSetBstMidEndSepPunct{\mcitedefaultmidpunct}
{\mcitedefaultendpunct}{\mcitedefaultseppunct}\relax
\EndOfBibitem
\bibitem{Bierlich:2022ned}
C.~Bierlich, S.~Chakraborty, G.~Gustafson, and L.~L\"onnblad, \ifthenelse{\boolean{articletitles}}{\emph{{Strangeness enhancement across collision systems without a plasma}}, }{}\href{https://doi.org/10.1016/j.physletb.2022.137571}{Phys.\ Lett.\  \textbf{B835} (2022) 137571}, \href{http://arxiv.org/abs/2205.11170}{{\normalfont\ttfamily arXiv:2205.11170}}\relax
\mciteBstWouldAddEndPuncttrue
\mciteSetBstMidEndSepPunct{\mcitedefaultmidpunct}
{\mcitedefaultendpunct}{\mcitedefaultseppunct}\relax
\EndOfBibitem
\bibitem{LHCb-DP-2008-001}
LHCb collaboration, A.~A. Alves~Jr.\ {\em et~al.}, \ifthenelse{\boolean{articletitles}}{\emph{{The \lhcb detector at the LHC}}, }{}\href{https://doi.org/10.1088/1748-0221/3/08/S08005}{JINST \textbf{3} (2008) S08005}\relax
\mciteBstWouldAddEndPuncttrue
\mciteSetBstMidEndSepPunct{\mcitedefaultmidpunct}
{\mcitedefaultendpunct}{\mcitedefaultseppunct}\relax
\EndOfBibitem
\bibitem{LHCb-DP-2014-002}
LHCb collaboration, R.~Aaij {\em et~al.}, \ifthenelse{\boolean{articletitles}}{\emph{{LHCb detector performance}}, }{}\href{https://doi.org/10.1142/S0217751X15300227}{Int.\ J.\ Mod.\ Phys.\  \textbf{A30} (2015) 1530022}, \href{http://arxiv.org/abs/1412.6352}{{\normalfont\ttfamily arXiv:1412.6352}}\relax
\mciteBstWouldAddEndPuncttrue
\mciteSetBstMidEndSepPunct{\mcitedefaultmidpunct}
{\mcitedefaultendpunct}{\mcitedefaultseppunct}\relax
\EndOfBibitem
\bibitem{LHCb-PAPER-2014-047}
LHCb collaboration, R.~Aaij {\em et~al.}, \ifthenelse{\boolean{articletitles}}{\emph{{Precision luminosity measurements at LHCb}}, }{}\href{https://doi.org/10.1088/1748-0221/9/12/P12005}{JINST \textbf{9} (2014) P12005}, \href{http://arxiv.org/abs/1410.0149}{{\normalfont\ttfamily arXiv:1410.0149}}\relax
\mciteBstWouldAddEndPuncttrue
\mciteSetBstMidEndSepPunct{\mcitedefaultmidpunct}
{\mcitedefaultendpunct}{\mcitedefaultseppunct}\relax
\EndOfBibitem
\bibitem{Sjostrand:2007gs}
T.~Sj\"{o}strand, S.~Mrenna, and P.~Skands, \ifthenelse{\boolean{articletitles}}{\emph{{A brief introduction to PYTHIA 8.1}}, }{}\href{https://doi.org/10.1016/j.cpc.2008.01.036}{Comput.\ Phys.\ Commun.\  \textbf{178} (2008) 852}, \href{http://arxiv.org/abs/0710.3820}{{\normalfont\ttfamily arXiv:0710.3820}}\relax
\mciteBstWouldAddEndPuncttrue
\mciteSetBstMidEndSepPunct{\mcitedefaultmidpunct}
{\mcitedefaultendpunct}{\mcitedefaultseppunct}\relax
\EndOfBibitem
\bibitem{LHCb-PROC-2010-056}
I.~Belyaev {\em et~al.}, \ifthenelse{\boolean{articletitles}}{\emph{{Handling of the generation of primary events in Gauss, the LHCb simulation framework}}, }{}\href{https://doi.org/10.1088/1742-6596/331/3/032047}{J.\ Phys.\ Conf.\ Ser.\  \textbf{331} (2011) 032047}\relax
\mciteBstWouldAddEndPuncttrue
\mciteSetBstMidEndSepPunct{\mcitedefaultmidpunct}
{\mcitedefaultendpunct}{\mcitedefaultseppunct}\relax
\EndOfBibitem
\bibitem{Pierog_2015}
T.~Pierog {\em et~al.}, \ifthenelse{\boolean{articletitles}}{\emph{{EPOS} {LHC}: Test of collective hadronization with data measured at the {CERN} {L}arge {H}adron {C}ollider}, }{}\href{https://doi.org/10.1103/physrevc.92.034906}{Phys.\ Rev.\  \textbf{C92} (2015) 034906}\relax
\mciteBstWouldAddEndPuncttrue
\mciteSetBstMidEndSepPunct{\mcitedefaultmidpunct}
{\mcitedefaultendpunct}{\mcitedefaultseppunct}\relax
\EndOfBibitem
\bibitem{Agostinelli:2002hh}
Geant4 collaboration, S.~Agostinelli {\em et~al.}, \ifthenelse{\boolean{articletitles}}{\emph{{Geant4: A simulation toolkit}}, }{}\href{https://doi.org/10.1016/S0168-9002(03)01368-8}{Nucl.\ Instrum.\ Meth.\  \textbf{A506} (2003) 250}\relax
\mciteBstWouldAddEndPuncttrue
\mciteSetBstMidEndSepPunct{\mcitedefaultmidpunct}
{\mcitedefaultendpunct}{\mcitedefaultseppunct}\relax
\EndOfBibitem
\bibitem{Allison:2006ve}
Geant4 collaboration, J.~Allison {\em et~al.}, \ifthenelse{\boolean{articletitles}}{\emph{{Geant4 developments and applications}}, }{}\href{https://doi.org/10.1109/TNS.2006.869826}{IEEE Trans.\ Nucl.\ Sci.\  \textbf{53} (2006) 270}\relax
\mciteBstWouldAddEndPuncttrue
\mciteSetBstMidEndSepPunct{\mcitedefaultmidpunct}
{\mcitedefaultendpunct}{\mcitedefaultseppunct}\relax
\EndOfBibitem
\bibitem{Clemencic:2011zza}
LHCb collaboration, M.~Clemencic {\em et~al.}, \ifthenelse{\boolean{articletitles}}{\emph{{The LHCb simulation application, Gauss: Design, evolution and experience}}, }{}\href{https://doi.org/10.1088/1742-6596/331/3/032023}{J.\ Phys.\ Conf.\ Ser.\  \textbf{331} (2011) 032023}\relax
\mciteBstWouldAddEndPuncttrue
\mciteSetBstMidEndSepPunct{\mcitedefaultmidpunct}
{\mcitedefaultendpunct}{\mcitedefaultseppunct}\relax
\EndOfBibitem
\bibitem{LHCb-DP-2012-004}
R.~Aaij {\em et~al.}, \ifthenelse{\boolean{articletitles}}{\emph{{The \lhcb trigger and its performance in 2011}}, }{}\href{https://doi.org/10.1088/1748-0221/8/04/P04022}{JINST \textbf{8} (2013) P04022}, \href{http://arxiv.org/abs/1211.3055}{{\normalfont\ttfamily arXiv:1211.3055}}\relax
\mciteBstWouldAddEndPuncttrue
\mciteSetBstMidEndSepPunct{\mcitedefaultmidpunct}
{\mcitedefaultendpunct}{\mcitedefaultseppunct}\relax
\EndOfBibitem
\bibitem{PDG2024}
Particle Data Group, S.~Navas {\em et~al.}, \ifthenelse{\boolean{articletitles}}{\emph{{\href{http://pdg.lbl.gov/}{Review of particle physics}}}, }{}\href{https://doi.org/10.1103/PhysRevD.110.030001}{Phys.\ Rev.\  \textbf{D110} (2024) 030001}\relax
\mciteBstWouldAddEndPuncttrue
\mciteSetBstMidEndSepPunct{\mcitedefaultmidpunct}
{\mcitedefaultendpunct}{\mcitedefaultseppunct}\relax
\EndOfBibitem
\bibitem{atomic_weight}
T.~Prohaska {\em et~al.}, \ifthenelse{\boolean{articletitles}}{\emph{{Standard atomic weights of the elements 2021 (IUPAC technical report)}}, }{}\href{https://doi.org/doi:10.1515/pac-2019-0603}{Pure and Applied Chemistry \textbf{94} (2022) 573}\relax
\mciteBstWouldAddEndPuncttrue
\mciteSetBstMidEndSepPunct{\mcitedefaultmidpunct}
{\mcitedefaultendpunct}{\mcitedefaultseppunct}\relax
\EndOfBibitem
\bibitem{LHCb-PAPER-2022-014}
LHCb collaboration, R.~Aaij {\em et~al.}, \ifthenelse{\boolean{articletitles}}{\emph{{Charmonium production in $\proton$Ne collisions at $\sqsnn = 68.5$ GeV }}, }{}\href{https://doi.org/10.1140/epjc/s10052-023-11608-6}{Eur.\ Phys.\ J.\  \textbf{C83} (2023) 625}, \href{http://arxiv.org/abs/2211.11645}{{\normalfont\ttfamily arXiv:2211.11645}}\relax
\mciteBstWouldAddEndPuncttrue
\mciteSetBstMidEndSepPunct{\mcitedefaultmidpunct}
{\mcitedefaultendpunct}{\mcitedefaultseppunct}\relax
\EndOfBibitem
\bibitem{LHCb-PAPER-2018-031}
LHCb collaboration, R.~Aaij {\em et~al.}, \ifthenelse{\boolean{articletitles}}{\emph{{Measurement of antiproton production in \proton{}He collisions at $\sqsnn=110$ GeV}}, }{}\href{https://doi.org/10.1103/PhysRevLett.121.222001}{Phys.\ Rev.\ Lett.\  \textbf{121} (2018) 222001}, \href{http://arxiv.org/abs/1808.06127}{{\normalfont\ttfamily arXiv:1808.06127}}\relax
\mciteBstWouldAddEndPuncttrue
\mciteSetBstMidEndSepPunct{\mcitedefaultmidpunct}
{\mcitedefaultendpunct}{\mcitedefaultseppunct}\relax
\EndOfBibitem
\bibitem{LHCb-DP-2013-002}
LHCb collaboration, R.~Aaij {\em et~al.}, \ifthenelse{\boolean{articletitles}}{\emph{{Measurement of the track reconstruction efficiency at LHCb}}, }{}\href{https://doi.org/10.1088/1748-0221/10/02/P02007}{JINST \textbf{10} (2015) P02007}, \href{http://arxiv.org/abs/1408.1251}{{\normalfont\ttfamily arXiv:1408.1251}}\relax
\mciteBstWouldAddEndPuncttrue
\mciteSetBstMidEndSepPunct{\mcitedefaultmidpunct}
{\mcitedefaultendpunct}{\mcitedefaultseppunct}\relax
\EndOfBibitem
\bibitem{Bierlich:2022pfr}
C.~Bierlich {\em et~al.}, \ifthenelse{\boolean{articletitles}}{\emph{{A comprehensive guide to the physics and usage of PYTHIA 8.3}}, }{}\href{https://doi.org/10.21468/SciPostPhysCodeb.8}{SciPost Phys.\ Codeb.\  \textbf{8} (2022) }, \href{http://arxiv.org/abs/2203.11601}{{\normalfont\ttfamily arXiv:2203.11601}}\relax
\mciteBstWouldAddEndPuncttrue
\mciteSetBstMidEndSepPunct{\mcitedefaultmidpunct}
{\mcitedefaultendpunct}{\mcitedefaultseppunct}\relax
\EndOfBibitem
\bibitem{Werner:2023zvo}
K.~Werner, \ifthenelse{\boolean{articletitles}}{\emph{{Revealing a deep connection between factorization and saturation: new insight into modeling high-energy proton-proton and nucleus-nucleus scattering in the EPOS4 framework}}, }{}\href{https://doi.org/10.1103/PhysRevC.108.064903}{Phys.\ Rev.\  \textbf{C108} (2023) 064903}, \href{http://arxiv.org/abs/2301.12517}{{\normalfont\ttfamily arXiv:2301.12517}}\relax
\mciteBstWouldAddEndPuncttrue
\mciteSetBstMidEndSepPunct{\mcitedefaultmidpunct}
{\mcitedefaultendpunct}{\mcitedefaultseppunct}\relax
\EndOfBibitem
\bibitem{Werner:2023fne}
K.~Werner and B.~Guiot, \ifthenelse{\boolean{articletitles}}{\emph{{Perturbative QCD concerning light and heavy flavor in the EPOS4 framework}}, }{}\href{https://doi.org/10.1103/PhysRevC.108.034904}{Phys.\ Rev.\  \textbf{C108} (2023) 034904}, \href{http://arxiv.org/abs/2306.02396}{{\normalfont\ttfamily arXiv:2306.02396}}\relax
\mciteBstWouldAddEndPuncttrue
\mciteSetBstMidEndSepPunct{\mcitedefaultmidpunct}
{\mcitedefaultendpunct}{\mcitedefaultseppunct}\relax
\EndOfBibitem
\bibitem{Werner:2023mod}
K.~Werner, \ifthenelse{\boolean{articletitles}}{\emph{{Parallel scattering, saturation, and generalized Abramovskii-Gribov-Kancheli (AGK) theorem in the EPOS4 framework, with applications for heavy-ion collisions at $\sqrt{s_{NN}}$ of 5.02\tev and 200\gev}}, }{}\href{https://doi.org/10.1103/PhysRevC.109.034918}{Phys.\ Rev.\  \textbf{C109} (2024) 034918}, \href{http://arxiv.org/abs/2310.09380}{{\normalfont\ttfamily arXiv:2310.09380}}\relax
\mciteBstWouldAddEndPuncttrue
\mciteSetBstMidEndSepPunct{\mcitedefaultmidpunct}
{\mcitedefaultendpunct}{\mcitedefaultseppunct}\relax
\EndOfBibitem
\bibitem{Werner:2023jps}
K.~Werner, \ifthenelse{\boolean{articletitles}}{\emph{{Core-corona procedure and microcanonical hadronization to understand strangeness enhancement in proton-proton and heavy ion collisions in the EPOS4 framework}}, }{}\href{https://doi.org/10.1103/PhysRevC.109.014910}{Phys.\ Rev.\  \textbf{C109} (2024) 014910}, \href{http://arxiv.org/abs/2306.10277}{{\normalfont\ttfamily arXiv:2306.10277}}\relax
\mciteBstWouldAddEndPuncttrue
\mciteSetBstMidEndSepPunct{\mcitedefaultmidpunct}
{\mcitedefaultendpunct}{\mcitedefaultseppunct}\relax
\EndOfBibitem
\bibitem{Bierlich:2018xfw}
C.~Bierlich, G.~Gustafson, L.~L\"onnblad, and H.~Shah, \ifthenelse{\boolean{articletitles}}{\emph{{The Angantyr model for heavy-ion collisions in PYTHIA8}}, }{}\href{https://doi.org/10.1007/JHEP10(2018)134}{JHEP \textbf{10} (2018) 134}, \href{http://arxiv.org/abs/1806.10820}{{\normalfont\ttfamily arXiv:1806.10820}}\relax
\mciteBstWouldAddEndPuncttrue
\mciteSetBstMidEndSepPunct{\mcitedefaultmidpunct}
{\mcitedefaultendpunct}{\mcitedefaultseppunct}\relax
\EndOfBibitem
\end{mcitethebibliography}

\newpage
% LHCb collaboration author list
% Data extracted on November 7th, 2024 at 3:29pm for paper reference LHCb-PAPER-2024-036
\centerline
{\large\bf LHCb collaboration}
\begin
{flushleft}
\small
R.~Aaij$^{38}$\lhcborcid{0000-0003-0533-1952},
A.S.W.~Abdelmotteleb$^{57}$\lhcborcid{0000-0001-7905-0542},
C.~Abellan~Beteta$^{51}$,
F.~Abudin{\'e}n$^{57}$\lhcborcid{0000-0002-6737-3528},
T.~Ackernley$^{61}$\lhcborcid{0000-0002-5951-3498},
A. A. ~Adefisoye$^{69}$\lhcborcid{0000-0003-2448-1550},
B.~Adeva$^{47}$\lhcborcid{0000-0001-9756-3712},
M.~Adinolfi$^{55}$\lhcborcid{0000-0002-1326-1264},
P.~Adlarson$^{82}$\lhcborcid{0000-0001-6280-3851},
C.~Agapopoulou$^{14}$\lhcborcid{0000-0002-2368-0147}
% ...

{\footnotesize \it

% ...
$^{14}$Universit{\'e} Paris-Saclay, CNRS/IN2P3, IJCLab, Orsay, France\\
% ...
$^{38}$Nikhef National Institute for Subatomic Physics, Amsterdam, Netherlands\\
% ...
$^{47}$Instituto Galego de F{\'\i}sica de Altas Enerx{\'\i}as (IGFAE), Universidade de Santiago de Compostela, Santiago de Compostela, Spain\\
% ...
$^{51}$Physik-Institut, Universit{\"a}t Z{\"u}rich, Z{\"u}rich, Switzerland\\
% ...
$^{55}$H.H. Wills Physics Laboratory, University of Bristol, Bristol, United Kingdom\\
% ...
$^{57}$Department of Physics, University of Warwick, Coventry, United Kingdom\\
% ...
$^{60}$School of Physics and Astronomy, University of Glasgow, Glasgow, United Kingdom\\
$^{61}$Oliver Lodge Laboratory, University of Liverpool, Liverpool, United Kingdom\\
% ...
$^{69}$Syracuse University, Syracuse, NY, United States\\
% ...
$^{82}$Department of Physics and Astronomy, Uppsala University, Uppsala, Sweden, associated to $^{60}$\\
% ...
}
\end{flushleft}

%The author list for journal publications is generated from the
%Membership Database shortly after 'approval to go to paper' has been
%given.  It is available at \url{https://lbfence.cern.ch/membership/authorship}
%and will be sent to you by email shortly after a paper number has been assigned.  
% The author list should be included in the draft used for  first and second circulation, to allow new members of the collaboration to verify that they have been included correctly. Occasionally a misspelled name is corrected, or associated institutions become full members. Therefore an updated author list will be sent to you after the final EB review of the paper.  In case line numbering doesn't work well after including the authorlist, try moving the \verb!\bigskip! after the last author to a separate line.

\end{document}